\documentclass[a4paper]{jpconf}
\usepackage{natbib}
\usepackage{graphicx}
\begin{document}
\title{Prospects for Detection of Extragalactic Stellar Black Hole Binaries in the Nearby Universe}

\author{Matthew Benacquista$^1$, Jesus Hinojosa$^1$, Alberto Mata$^1$, Krzysztof Belczynski$^{1,2}$ }

\address{$^1$ Center for Gravitational Wave Astronomy, University of Texas at Brownsville, One University Blvd, Brownsville, TX 78520}
\address{$^2$ Astronomical Observatory, University of Warsaw, Al. Ujazdowskie 4, 00-478 Warsaw, Poland}

\ead{matthew.benacquista@utb.edu}

\begin{abstract}
Stellar mass black hole binaries have individual masses between 10-80 solar masses. These systems may emit gravitational waves at frequencies detectable at Megaparsec distances by space-based gravitational wave observatories. In a previous study, we determined the selection effects of observing these systems with detectors similar to the Laser Interferometer Space Antenna by using a generated population of binary black holes that covered a reasonable parameter space and calculating their signal-to-noise ratio. We further our study by populating the galaxies in our nearby (less than 30 Mpc) universe with binary black hole systems drawn from a distribution found in the Synthetic Universe to ultimately investigate the likely event rate of detectable binaries  from galaxies in the nearby universe.
\end{abstract}

\section{Introduction}
Space-based gravitational wave interferometers of the LISA design will be sensitive to mHz gravitational waves. The latest design for eLISA consists of a two-arm interferometer with 1 million km arm lengths~\cite{gravitationaluniverse13}. The majority of stellar mass compact object binary sources for eLISA will be the Galactic population of close white dwarf binaries. With a typical chirp mass of $\sim 0.5~{\rm M_\odot}$, white dwarf binaries are unlikely to be detected from extragalactic distances since their signal strength will be too low. It is possible that a few binary black holes in the Milky Way may also be in the eLISA band~\cite{belczynski10a}. These systems may have much higher chirp masses (up to $\sim 60~{\rm M_\odot}$) and may therefore be detectable at Mpc distances.

In order to find a first estimate of the likelihood of observing extragalactic stellar mass binary black holes with eLISA, we have performed a population synthesis of these systems within the local universe out to 30 Mpc. We have used the Gravitational Wave Galaxy Catalogue to obtain the spatial distribution of host galaxies within this range. We then populated the galaxies using a binary black hole population synthesis available from the Synthetic Universe. Finally, we estimate the probability of detection for each binary within the eLISA band.

\section{Population Synthesis}
The Gravitational Wave Galaxy Catalogue (GWGC) was put together in order to assist in electromagnetic follow-up searches for gravitational wave triggers generated by ground-based detectors such as LIGO and Virgo~\cite{white11}. These triggers were expected to arise from the inspirals of binary black holes and neutron stars. The catalogue is ``a more complete, up-to-date sample created from a variety of literature sources extending out to 100 Mpc, which is as unbiased as possible to a particular kind of source.''~\cite{white11} The GWGC contains sky position, distance, blue magnitude, major and minor diameters, position angle, and galaxy type for 53,225 galaxies out to 100 Mpc~\cite{white12}. We have made a distance cut to galaxies within 30 Mpc. This results in a population of 8,947 galaxies to populate with binary black holes.

In order to populate the galaxies with binary black holes, we grouped the galaxies in the catalog into three basic galaxy types: Spiral, Elliptical, and Irregular. We populated these galaxies using a population synthesis of binary black holes generated by Belczynski and collaborators using StarTrack and publicly available at {\tt SyntheticUniverse.org}. There are 32,496 binary black holes in this population synthesis. In order to determine the number of binaries to include in each galaxy, we assume that the blue luminosity correlates to the total binary black hole population. Based upon a population synthesis for the Milky Way which gives roughly $10^6$ binary black holes~\cite{belczynski10a}, we have assigned $10^6$ binary black holes for every Milky Way equivalent blue luminosity for each galaxy in the catalogue.

Once we have determined the number of binary black holes to assign to each galaxy, we then select a star formation history for each galaxy based on its morphological type. We assume that spiral galaxies have had a constant star formation rate over the past 10 Gyr. Elliptical galaxies are given a burst of star formation between 9 and 10 Gyr, while irregular galaxies have been assigned a recent burst of star formation over the past 3 Gyr.

Based on these star formation rates, we then populate each galaxy by randomly selecting a binary from the pool of binary black holes taken from the synthetic universe and giving it a random age within the bounds governed by the star formation history appropriate to its morphological type. We repeat this process until we have selected the number of black holes expected by the blue luminosity. The orbital period of each binary is then propagated forward to the present day under the influence of the emission of gravitational radiation. We retain only those binaries with a final orbital period below 2000 s. These systems will be at gravitational wave frequencies above 1 mHz and therefore they will lie above the confusion-limited noise from the Galactic population of close white dwarf binaries. This resulted in 45,112 potential eLISA sources in galaxies within 30 Mpc. The populations of each galaxy type are shown in Table~\ref{tab:potentialsources}.
\begin{table}[h]
\caption{\label{tab:potentialsources} Number of galaxies of each morphological type within 30 Mpc and the number of binary black holes that are potential eLISA sources within each type.}
\begin{center}
\begin{tabular}{lrr}
Galaxy & Number of & Number of\\
type &  galaxies & binaries \\
\mr
Elliptical & 689 & 708\\
Spiral & 7137 & 22128\\
Irregular & 1121 & 22276\\
\br
\end{tabular}
\end{center}
\end{table}

\section{Detection Likelihood}
\begin{figure}[h]
\includegraphics[width=\textwidth]{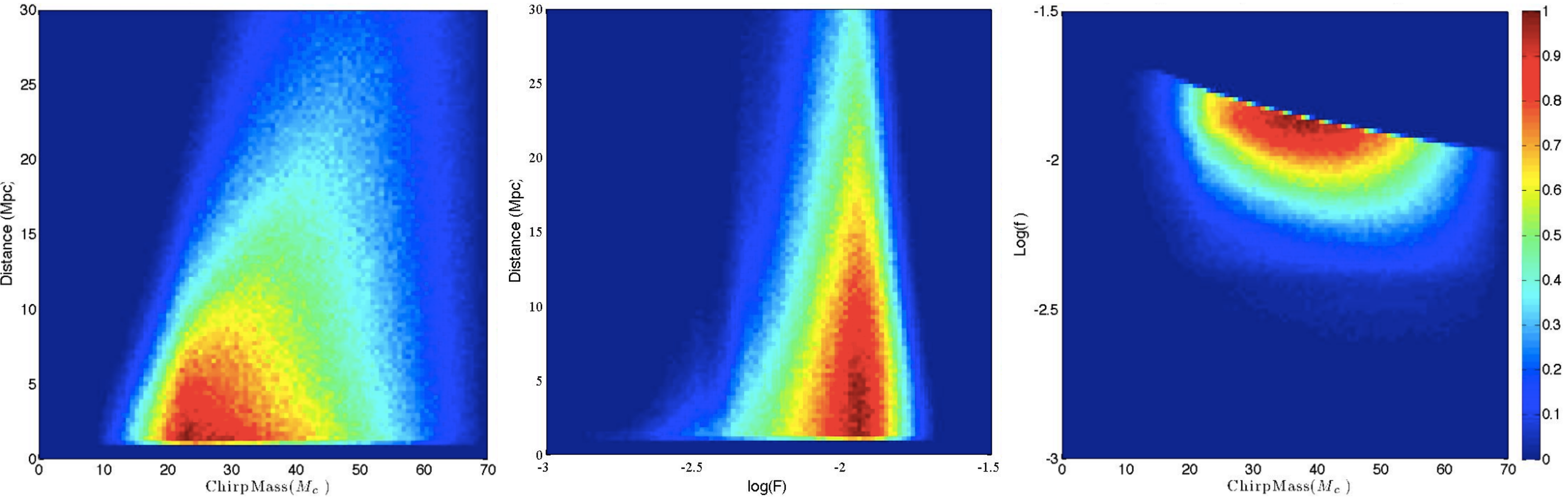}
\caption{\label{fig:likelihoods}Likelihood of detection for binary black holes obtained from the Monte Carlo study. The likelihoods have been marginalized over frequency on the left, chirp mass in the center, and distance on the right. The abrupt cutoff in the likelihoods on the right is due to systems that evolve out of the eLISA band in one year.} 
\end{figure}

Once we have determined the population of potential eLISA sources, we need to determine the likelihood that these binaries will be detectable by eLISA. As this is a preliminary study, we haven't done multiple realizations of the population. Instead, we have run a Monte Carlo study of $10^9$ binaries with an even coverage over the parameter space of black hole masses, orbital period, and distance. The black hole masses cover the range from $M \in \left(6,80\right)~{\rm M_\odot}$, the orbital periods cover the range $P_{\rm orb} \in \left(0,2000\right)~{\rm s}$, and the distances are less than 30 Mpc. The orbital inclinations and sky locations were randomly chosen. All of these binaries were then run through an eLISA simulator to determine the expected signal to noise ratio. We include the gravitational chirp in the evolution of this sample of binaries. Only those binaries that remain within the eLISA band for a full year are included in the signal to noise ratio calculations. Using the initial pool from the Monte Carlo study and the pool of binaries with a signal to noise ratio above the threshold, we can determine the likelihood of detection for any binary with a given chirp mass, orbital period, and distance. Projections of the likelihood are shown in Figure~\ref{fig:likelihoods}.

Using the known locations of the populated galaxies in the catalogue along with the binary properties to the potential eLISA sources, we can then estimate the likelihood of detection for each binary in the population. Although this approach averages over orbital inclinations and doesn't take into account the sky location of each host galaxy, we can obtain a first order estimate of the number of observable extragalactic binaries. There were no observable binaries within the elliptical galaxy population, five potentially observable binaries in the spiral galaxy population, and five in the irregular galaxy population. Out of the ten, two of these binaries had a greater than 50\% probability of detection. The results are shown in Table~\ref{tab:detectionprob}
\begin{table}[h]
\caption{\label{tab:detectionprob}Probability for detection from the 10 binary black holes.}
\begin{center}
\begin{tabular}{lr@{.}lr@{.}lr@{.}lr@{.}lr@{.}l}
Galaxy type & \multicolumn{10}{c}{Detection Probability (\%)}\\
\mr
Spiral & 89 & 6 & 2 & 2 & 2 & 2 & 1 & 7 & 0 & 3 \\
Irregular & 51 & 1 & 17 & 2 & 9 & 7 & 2 & 1 & 2 & 1\\
\br
\end{tabular}
\end{center}
\end{table}

\section{Conclusions}

We have used a very simple population synthesis technique to perform an initial investigation into the possibility of observing stellar mass binary black holes at extragalactic distances. For one realization of the population, using simple prescriptions for the population as a function of the blue luminosity and ignoring metallicity and specific star formation rates for individual galaxies, we have estimated the likelihood of observing an extragalactic binary black hole with eLISA. The results from this single realization indicate one or two binary black holes will be observed to within 30 Mpc, although there is a non-zero probability of up to 10 binary black holes being observed.

Our simulation did not include the population of black hole binaries that may arise in the globular cluster systems of the galaxies within the Gravitational Wave Galaxy Catalog. Globular cluster systems will likely come from lower metallicity populations and therefore may contain more high mass black holes which will enhance the number of observable black hole binaries. These results are promising enough to warrant further study into the population of black hole binaries within the local universe. 

\ack
This work was funded by the Center of Gravitational Wave Astronomy, NSF and NASA, through grants NASA NNX09AV06A and NSF HRD0734800. JH and AM also acknowledge the support of the Arecibo Remote Command Center scholars program, funded by NSF AST0750913.

\bibliography{HinojosaBenacquista}

\end{document}